\documentclass[twocolumn]{aastex7}

\usepackage{amsmath}

\received{2025 March 18}
\revised{2025 May 1}
\accepted{2025 May 10}
\submitjournal{ApJ Letters}

\shorttitle{Spectropolarimetry of A Nuclear Transient AT2023clx}
\shortauthors{Uno et al.}

\graphicspath{{./}{figures/}}

\begin{document}

\title{Spectropolarimetry of A Nuclear Transient AT2023clx: Revealing The Geometrical Alignment between The Transient Outflow and The Nuclear Dusty Region}

\author[0000-0002-6765-8988]{Kohki Uno}
\affiliation{Department of Astronomy, Kyoto University, Kitashirakawa-Oiwake-cho, Sakyo-ku, Kyoto, 606-8502, Japan}
\email[show]{k.uno@kusastro.kyoto-u.ac.jp} 

\author[0000-0003-2611-7269]{Keiichi Maeda}
\affiliation{Department of Astronomy, Kyoto University, Kitashirakawa-Oiwake-cho, Sakyo-ku, Kyoto, 606-8502, Japan}
\email{keiichi.maeda@kusastro.kyoto-u.ac.jp}

\author[0000-0002-3933-7861]{Takashi Nagao}
\affiliation{Department of Physics and Astronomy, University of Turku, FI-20014 Turku, Finland}
\affiliation{Aalto University Metsähovi Radio Observatory, Metsähovintie 114, 02540 Kylmälä, Finland}
\affiliation{Aalto University Department of Electronics and Nanoengineering, P.O. BOX 15500, FI-00076 AALTO, Finland}
\email{takashi.nagao@utu.fi}

\author[0000-0002-8597-0756]{Giorgos Leloudas}
\affiliation{DTU Space, National Space Institute, Technical University of Denmark, Elektrovej 327, 2800 Kgs. Lyngby, Denmark}
\email{giorgos@space.dtu.dk}

\author[0000-0002-0326-6715]{Panos Charalampopoulos}
\affiliation{Department of Physics and Astronomy, University of Turku, FI-20014 Turku, Finland}
\email{pachar@utu.fi}

\author[0000-0001-7497-2994]{Seppo Mattila}
\affiliation{Department of Physics and Astronomy, University of Turku, FI-20014 Turku, Finland}
\affiliation{School of Sciences, European University Cyprus, Diogenes Street, Engomi, 1516, Nicosia, Cyprus}
\email{sepmat@utu.fi}

\author[0000-0003-4569-1098]{Kentaro Aoki}
\affiliation{Subaru Telescope, National Astronomical Observatory of Japan, 650 North A’ohoku Place, Hilo, HI 96720, USA}
\email{kaoki@naoj.org}

\author[0000-0002-8482-8993]{Kenta Taguchi}
\affiliation{Department of Astronomy, Kyoto University, Kitashirakawa-Oiwake-cho, Sakyo-ku, Kyoto, 606-8502, Japan}
\affiliation{Okayama Observatory, Kyoto University, 3037-5 Honjo, Kamogatacho, Asakuchi, Okayama 719-0232, Japan}
\email{kentagch@kusastro.kyoto-u.ac.jp}

\author[0000-0002-4540-4928]{Miho Kawabata}
\affiliation{Okayama Observatory, Kyoto University, 3037-5 Honjo, Kamogatacho, Asakuchi, Okayama 719-0232, Japan}
\email{kawabata@kusastro.kyoto-u.ac.jp}

\author[0000-0002-8079-7608]{Javier Moldon}
\affiliation{Instituto de Astrof\'isica de Andaluc\'ia (IAA, CSIC), Glorieta de las Astronom\'ia, s/n, E-18008 Granada, Spain}
\email{jmoldon@iaa.es}

\author[0000-0001-5654-0266]{Miguel P\'erez-Torres}
\affiliation{Instituto de Astrof\'isica de Andaluc\'ia (IAA, CSIC), Glorieta de las Astronom\'ia, s/n, E-18008 Granada, Spain}
\affiliation{School of Sciences, European University Cyprus, Diogenes Street, Engomi, 1516, Nicosia, Cyprus}
\email{torres@iaa.es}

\author[0000-0003-4663-4300]{Miika Pursiainen}
\affiliation{Department of Physics, University of Warwick, Gibbet Hill Road, Coventry, CV4 7AL, UK}
\email{miika.pursiainen@warwick.ac.uk}

\author[0000-0002-1022-6463]{Thomas Reynolds}
\affiliation{Department of Physics and Astronomy, University of Turku, FI-20014 Turku, Finland}
\affiliation{Cosmic Dawn Center (DAWN)}
\affiliation{Niels Bohr Institute, University of Copenhagen, Jagtvej 128, 2200 København N, Denmark}
\email{treynolds1729@gmail.com}

\begin{abstract}

AT2023clx, which occurred in NGC3799 with a Low-Ionization Nuclear Emission-Line Region (LINER), is one of the most nearby nuclear transients classified as a tidal disruption event (TDE). We present three-epoch spectropolarimetric follow-up observations of AT2023clx. We detected two polarization components; one is a constant polarization of $\sim 1\%$ originated from an aspherical outflow associated with the transient, while the other is a blue-excess polarization toward $\sim 2\%$ originated from a nuclear dusty environment via light echoes. The polarization angle flipped by 90 degrees between the two epochs, indicating that the outflow direction was perpendicular to the dust plane. Furthermore, the polarized flux might suggest that the nuclear dust favors relatively large grains, potentially offering constraints on its physical properties. Such polarization features -- the blue excess and the 90-degree flip -- have never been observed in previous TDE polarization samples, highlighting unique mechanisms behind AT2023clx. We propose possible scenarios: the disruption of a star formed within or captured by a nuclear dusty cloud. Given the LINER nature of NGC3799, the dusty region may possibly be linked to a torus or disk associated with a weak Active Galactic Nucleus (AGN). Furthermore, as a more speculative scenario, the event might have been triggered by AGN-like activity, potentially linked to changing-look AGNs or ambiguous nuclear transients. These findings highlight the power of time-series spectropolarimetry of TDEs, not only in probing the origins of nuclear transients, but also in investigating the physical properties of nuclear dust.
\end{abstract}

\keywords{
\uat{Tidal disruption}{1696} --- 
\uat{Polarimetry}{1278} --- 
\uat{Spectropolarimetry}{1973} --- 
\uat{Galaxy nuclei}{609}
}

\section{Introduction} \label{sec:intro}

Supermassive black holes (SMBHs) located at the centers of galaxies are key drivers of time-dependent astrophysical phenomena called nuclear transients, with notable examples being outbursts of active galactic nuclei \citep[AGNs;][]{Antonucci1993ARAA, Urry1995PASP} and tidal disruption events \citep[TDEs;][]{Rees1988Nature}. AGNs are luminous galaxy cores powered by the mass accretion onto their SMBHs through surrounding accretion disks in dust-rich environments, and some of them exhibit luminous flares. On the other hand, TDEs are astrophysical phenomena where stars are dynamically torn apart by the tidal forces of SMBHs, resulting in luminous transients in multi-wavelength covering X-ray \citep{Bade1996AA}, radio \citep{Bloom2011Sci}, optical/ ultraviolet \citep[UV,][]{vanVelzen2011ApJ, vanVelzen2021ApJ}, and infrared \citep{Mattila2018Sci, Masterson2024ApJ}. The typical timescale of the luminous flares in AGNs is comparable to or longer than that of TDEs \citep{Yang2018ApJ, Frederick2019ApJ, vanVelzen2021ApJ}, and the spectral properties of AGNs are typically distinct from those of TDEs. However, the link between the observational properties and their physical origins has not been established in full detail; it currently relies largely on phenomenological grounds. Indeed, some overlaps between the origins of AGNs and TDEs are inferred from an emerging group of `ambiguous nuclear transients' \citep[ANTs;][]{Hinkle2022ApJ} that exhibit both TDE-like and AGN-like characteristics and tend to occur in weak AGNs. These phenomena may suggest that a fraction of previously observed nuclear transients that exhibit TDE signatures may require some conditions beyond, or even have a different origin from, classical TDEs. The nuclear transient zoo may still include previously unknown populations bridging across different classes.

\begin{figure}[t]
\epsscale{1.12}
\plotone{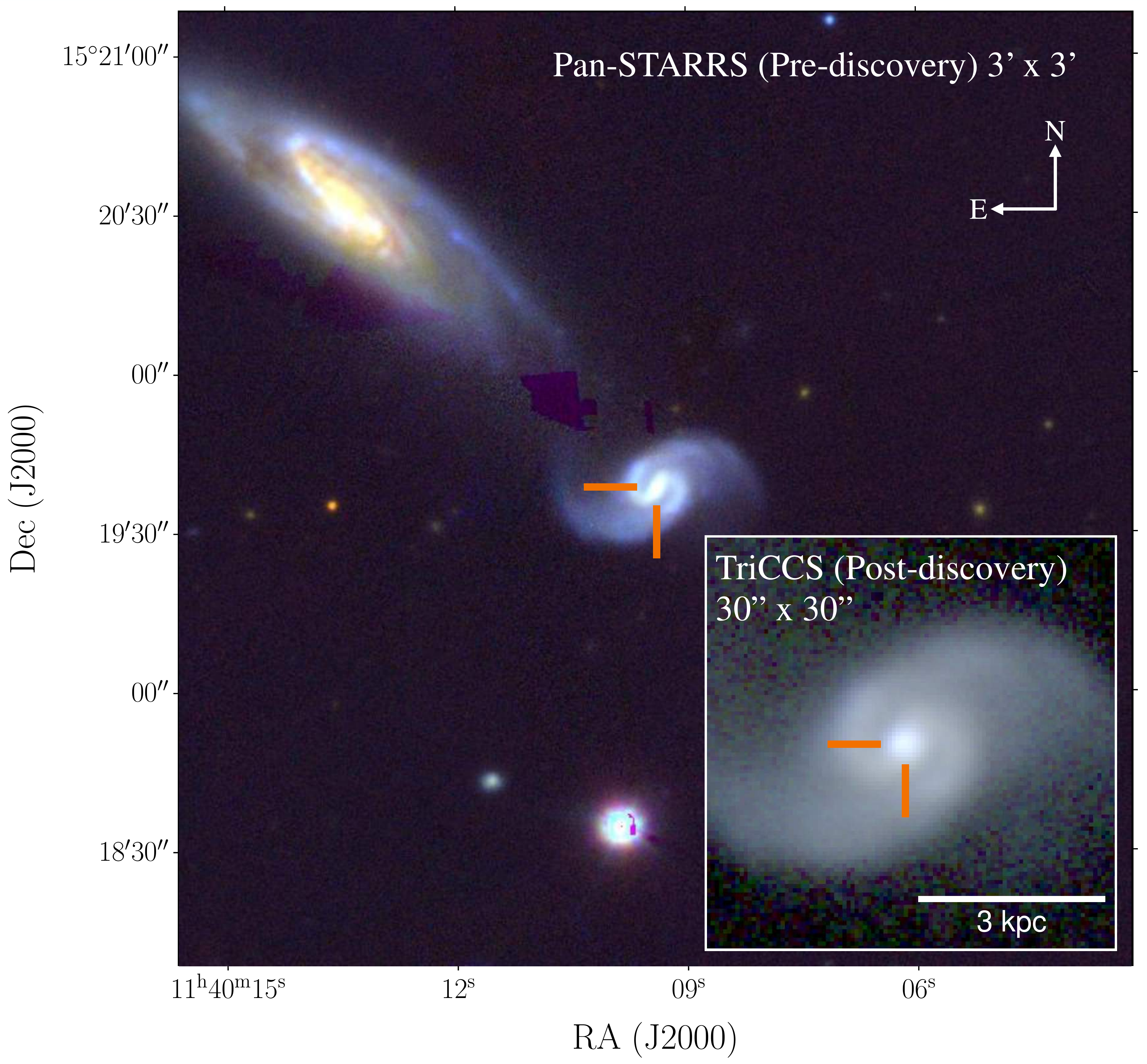}
\caption{
Pre-outburst image of the host galaxy of AT2023clx (NGC3799), created by combining $g$-, $i$-, and $y$-band images from the Pan-STARRS1 survey \citep{Kaiser2002SPIE}. The inset panel (bottom-right) shows an enlarged view of a post-outburst image, as obtained by combining the $g$-, $r$-, and $i$-band images taken by Seimei/TriCCS, showing AT2023clx on the core of NGC3799. NGC3799 is in an interacting system with NGC3800 (another galaxy seen in the top-left corner). 
}
\label{fig:finding}
\end{figure}

On February 22, 2023 (MJD 59997.21), AT2023clx was discovered by the ASAS-SN \citep{Stanek2023TNSTR}, and then it was classified as a TDE \citep{Taguchi2023TNSCR}. The transient coordinate was $\rm{RA(J2000.0)} = 11^{\rm h}40^{\rm m}09^{\rm s}.40$ and ${\rm Dec(J2000.0)} = 15^{\circ}19^{\prime}38^{\prime\prime}.5$, coincided with the geometric centroid of the host galaxy (NGC3799) at the redshift of $z=0.01107$ (Fig. \ref{fig:finding}). NGC3799 is a spiral galaxy, which has been classified as a Type II Low-Ionization Nuclear Emission-Line Region \cite[LINER;][]{Heckman1980AA} with the central BH mass of $\log\left(M_{\rm BH} / M_{\odot}\right) = 6.26\pm0.28$ \citep{Zaw2019ApJ}. Given the luminosity distance of $\sim 50{\rm ~Mpc}$, AT2023clx is one of the most nearby objects that have been classified as TDEs \citep{Zhu2023ApJL, Hoogendam2024MNRAS, Charalampopoulos2024AA}. Furthermore, the post-peak light curves followed the $t^{-5/3}$ decay \citep{Zhu2023ApJL}, which is also a characteristic property of TDEs \citep{Strubbe2009MNRAS}. The $g-$band peak magnitude of $M_{\rm g}=-17.16$ mag \citep{Zhu2023ApJL} suggests that AT2023clx is on the faintest end among optical/UV TDEs \citep[e.g.,][]{Blagorodnova2017ApJ, Mockler2019ApJ, Charalampopoulos2023AA}.

Immediately after the classification, we initiated a follow-up campaign for AT2023clx, and recorded the time evolution of polarization that has never been seen in previous TDE samples. This letter presents the results of our spectropolarimetric follow-up observations. In Section \ref{sec:observations}, we summarize our observations and data reduction. In Section \ref{sec:results}, we show the results of spectropolarimetry. In Section \ref{sec:discussion}, we show our interpretation of the time evolution of polarization, followed by discussion about the origin of AT2023clx. This letter is closed in Section \ref{sec:conclusions} with conclusions. In addition to the polarimetric observation, we also performed radio follow-up observations. We present the results in Appendix \ref{appendix:radio}.

\section{Observations and Data Reduction} \label{sec:observations}

\subsection{Follow-up Observations: Spectroscopy}

On February 26 ($+4.4$ days after the first detection), we obtained the classification spectrum of AT2023clx with KOOLS-IFU \citep{Matsubayashi2019PASJ} mounted on the Seimei telescope \citep{Kurita2020PASJ}, and reported the spectrum to the Transient Name Server \citep{Taguchi2023TNSCR}. An additional spectrum was also obtained with the Seimei telescope on March 3 ($+8.5$ days). We used the VPH-blue grism, covering $4100-8900$ \text{\AA}. The wavelength resolution ($R = \lambda/\Delta\lambda$) of VPH-blue was $\sim 500$. For the wavelength calibration, we used Hg, Ne, and Xe lamps. The data reduction was performed with the Hydra package in IRAF\footnote{IRAF is distributed by the National Optical Astronomy Observatory, which is operated by the Association of Universities for Research in Astronomy (AURA) under a cooperative agreement with the National Science Foundation.} and a set of self-developed routines that specifically handle the reduction of the KOOL-IFU data\footnote{\url{http://www.o.kwasan.kyoto-u.ac.jp/inst/p-kools/reduction-201806/index.html}}. In addition, we also obtained the post-outburst images with the TriColor CMOS Camera and Spectrograph (TriCCS) mounted on the Seimei telescope\footnote{\url{http://www.o.kwasan.kyoto-u.ac.jp/inst/triccs/}}. The observing log is shown in Table \ref{tab:observation}. 

\begin{deluxetable*}{ccccccccc}
\tablenum{1}
\tablecaption{Log of observations of AT2023clx\label{tab:observation}}
\tablewidth{0pt}
\tabletypesize{\scriptsize}
\tablehead{
\colhead{Date} & \colhead{MJD} & \colhead{Phase} & \colhead{Telescope} & \colhead{Mode} & \colhead{Filter/Grism} & \colhead{Exposure Time} & \colhead{Polarization ($V-$band)} & \colhead{Polarization ($R-$band)} \\
\nocolhead{} & \nocolhead{} & \colhead{[day]} & \colhead{} & \nocolhead{} & \nocolhead{} & \nocolhead{} & \colhead{(Q,U) [\%]} & \colhead{(Q,U) [\%]}
}
\startdata
  2023-02-26 & 60001.58 & 4.4  & Seimei & Spectroscopy    & VPH-blue & 900sec $\times$ 4sets & - & - \\
  2023-02-26 & 60001.62 & 4.4  & Seimei & Imaging       & g/r/i  & 120sec  & - & - \\
  2023-02-27 & 60002.57 & 5.4  & Subaru & Spectropolarimetry & B300   & 600sec $\times$ 2sets & $(-0.43, -0.66)$ & $(-0.17, -0.50)$\\
  2023-02-28 & 60003.04 & 5.9  & NOT  & Imaging Polarimetry & V    & 200sec $\times$ 1set & $(-0.69 \pm 0.21, -0.44 \pm 0.40)$ & - \\
  2023-03-02 & 60005.71 & 8.5  & Seimei & Spectroscopy    & VPH-blue & 600sec $\times$ 2sets & - & - \\
  2023-03-14 & 60017.20 & 20.0 & NOT  & Imaging Polarimetry & V    & 250sec $\times$ 1set & $(0.28 \pm 0.53, 1.72 \pm 0.31)$ & - \\
  2023-04-15 & 60049.26 & 52.1 & Subaru & Spectropolarimetry & B300   & 1200sec $\times$ 3sets & $(0.54, 0.98)$ & $(0.34, 0.72)$\\
  2023-04-15 & 60049.43 & 52.2 & Subaru & Imaging       & V/R   & 30sec  & - & - \\
  2023-06-22 & 60118.28 & 121.1 & Subaru & Spectropolarimetry & B300   & 600sec $\times$ 2sets & $(-0.15,  0.54)$ & $(0.01, 0.48)$\\
\enddata
\tablecomments{The phase is measured relative to the epoch of the first detection (MJD 59997.21). For polarimetry, the exposure time shown here is per a HWP angle. The polarization values of spectropolarimetry represent synthesized broadband polarizations, calculated by assuming the filter transmission curves of the NOT $V$-band $R$-bands. The associated uncertainties are sufficiently small ($\lesssim 0.001\%$). The polarization values of imaging polarimetry are obtained by averaging over different aperture sizes (see Appendix \ref{appendix:correction}).}
\end{deluxetable*}

\subsection{Follow-up Observations: Spectropolarimetry}

 Spectropolarimetry of AT2023clx was conducted at three epochs, February 27 ($+5.4$ days), April 15 ($+52.1$ days), and June 22 ($+121.1$ days). The observations were carried out using the Faint Object Camera and Spectrograph \citep[FOCAS;][]{Kashikawa2002PASJ, Kawabata2003SPIE} mounted on the Subaru telescope. We used the B300 grating with $1.0^{\prime\prime}$ center-slit and no order-sorting filter. While the second-order scattering light may (slightly) contaminate the signal above $\sim 7000$ {\text \AA}, we note that any of our conclusions would not rely on the data in this particular wavelength range. Under this configuration, the wavelength coverage is $3650-8300$ {\text \AA} with a spectral resolution of $R \sim 500$. We reduced the data with IRAF using the standard procedure for spectropolarimetry \citep[for details, see][]{Patat2017hsn}. We calibrated the wavelength using Th and Ar lamps.

FOCAS has a Wollaston prism and a rotating half-wave plate (HWP). The Wollaston prism splits the incident ray into two beams with orthogonal polarization directions, namely ordinary and extraordinary beams. Our spectropolarimetric data with FOCAS are composed of four frames for one set, corresponding to the HWP rotation angles of $0^{\circ}$, $22.5^{\circ}$, $45^{\circ}$, and $67.5^{\circ}$, i.e., $I_{0}$, $I_{45}$, $I_{90}$, and $I_{135}$, where $I_{\phi}$ is the flux where $\phi$ corresponds to twice the HWP angle. We define the Stokes parameters $Q$, $U$, and $P$, and the polarization angle $\theta$, as follows: $Q = (I_{0} - I_{90}) / (I_{0} + I_{90}) = (I_{0} - I_{90}) / I = P\cos 2\theta$, $U = (I_{45} - I_{135}) / (I_{45} + I_{135}) = (I_{45} - I_{135}) / I = P\sin 2\theta$, $P = \sqrt{Q^{2} + U^{2}}$, and $\theta = 1/2 \arctan\left( U/Q \right)$, where $I = I_{0} + I_{90} = I_{45} + I_{135}$ is the total flux (see Appendix \ref{appendix:correction} for more details).

\subsection{Follow-up Observations: Imaging Polarimetry}

We obtained the $V$-band imaging polarimetry using ALFOSC mounted on the Nordic Optical telescope (NOT) on February 28 ($+5.9$ days) and March 14 ($+20.0$ days). We analyzed the ordinary and extraordinary beams, and evaluated the Stokes parameters following a similar procedure to the spectropolarimetry \citep{Patat2017hsn}. Aperture photometry was performed in both the ordinary and extraordinary beams. Since the transient coincides with the nucleus of the host galaxy, the measurements were affected by the background light. A smaller aperture will reduce the background contamination, but will also lead to an incomplete integration of the emission from the nuclear transient source of interest. To estimate the error by the background, we used two aperture sizes; one is equal to the full width at half maximum (FWHM) of the ordinary beam’s point-spread function and the other is twice the size of the FWHM. We also defined two sky regions with inner radii of two or three times the size of the FWHM, and with a width equal to the FWHM. The mean and the error of the Stokes parameters derived from these two aperture sizes were assigned to be the measured values and the associated uncertainties, respectively. The latter also includes the photon shot noise.

\section{Results} \label{sec:results}

\subsection{Spectroscopy}

\begin{figure*}[t]
\epsscale{1.12}
\plotone{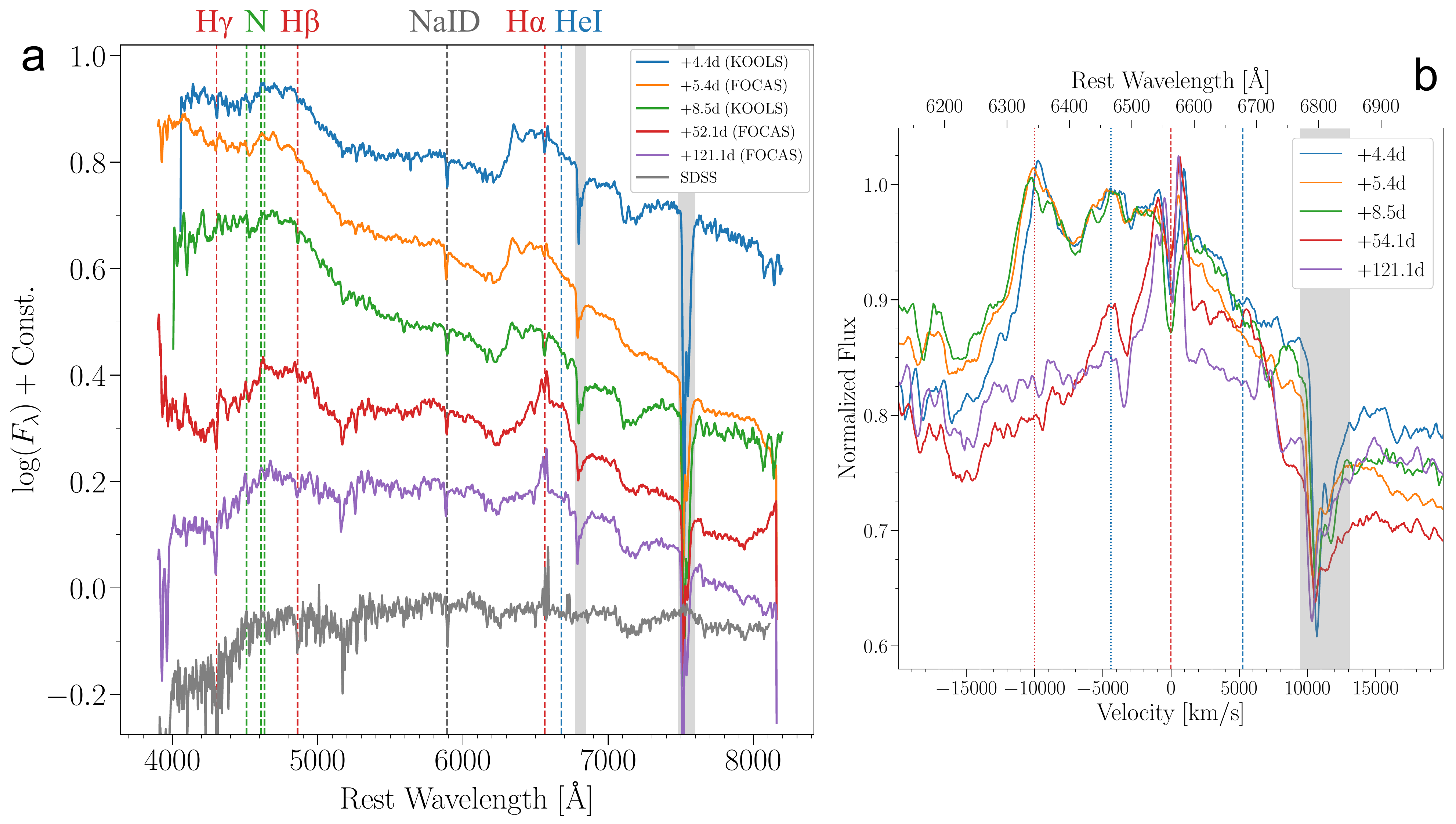}
\caption{
Left (Panel a): Time evolution of the spectra of AT2023clx from $+4$ days to $+121$ days after its first detection. Additionally, the pre-outburst spectrum at the position of AT2023clx taken by the Sloan Digital Sky Survey (SDSS) is presented by a gray line. All the spectra are shown in the rest wavelength, corrected for its redshift ($z = 0.01107$). The vertical dashed lines in red, blue, green, and gray correspond to the rest wavelengths of hydrogen, helium, nitrogen, and sodium lines, respectively. The gray-shaded regions are affected by the telluric absorption. In the early phases, the spectra displayed a blue continuum and broad hydrogen lines. In the later phases, the broad features became weaker. The overall spectral features on $+121$ days closely resembled the pre-outburst spectrum.
Right (Panel b): Time evolution of the $\rm H\alpha$ profile, plotted in velocity space. The rest wavelengths are indicated on the top-horizontal axis. The dashed lines show the rest wavelengths of hydrogen and helium lines. The dotted lines show the same atomic lines but with an additional blueshift of $10,000 {\rm ~km~s^{-1}}$. The gray-shaded region shows the wavelength range affected by the telluric absorption. The high-velocity component was diminishing toward later phases.
}
\label{fig:specevolution}
\end{figure*}

Fig. \ref{fig:specevolution}a shows the spectral evolution of AT2023clx. The first spectrum showed a blue continuum and broad Balmer lines. These spectral properties are typical of optical/UV TDEs \citep{Arcavi2014ApJ, vanVelzen2021ApJ, Charalampopoulos2022AA}. The spectra became redder as time went on, and the hydrogen lines gradually became narrower. The temporal evolution of the spectra suggests that the outburst component prevailed in the early spectra, while the host galaxy and LINER emission became increasingly dominant in the later phase. A striking feature is the presence of blue-shifted and distinct emission components with velocities of $\sim 10,000 {\rm ~km~s^{-1}}$ in the early hydrogen and helium lines (Fig. \ref{fig:specevolution}b), which faded away in the late phase \citep[see also,][]{Charalampopoulos2024AA}. 

\subsection{Polarimetry}

\subsubsection{Polarization Evolution}

\begin{figure*}[t]
\epsscale{1.12}
\plotone{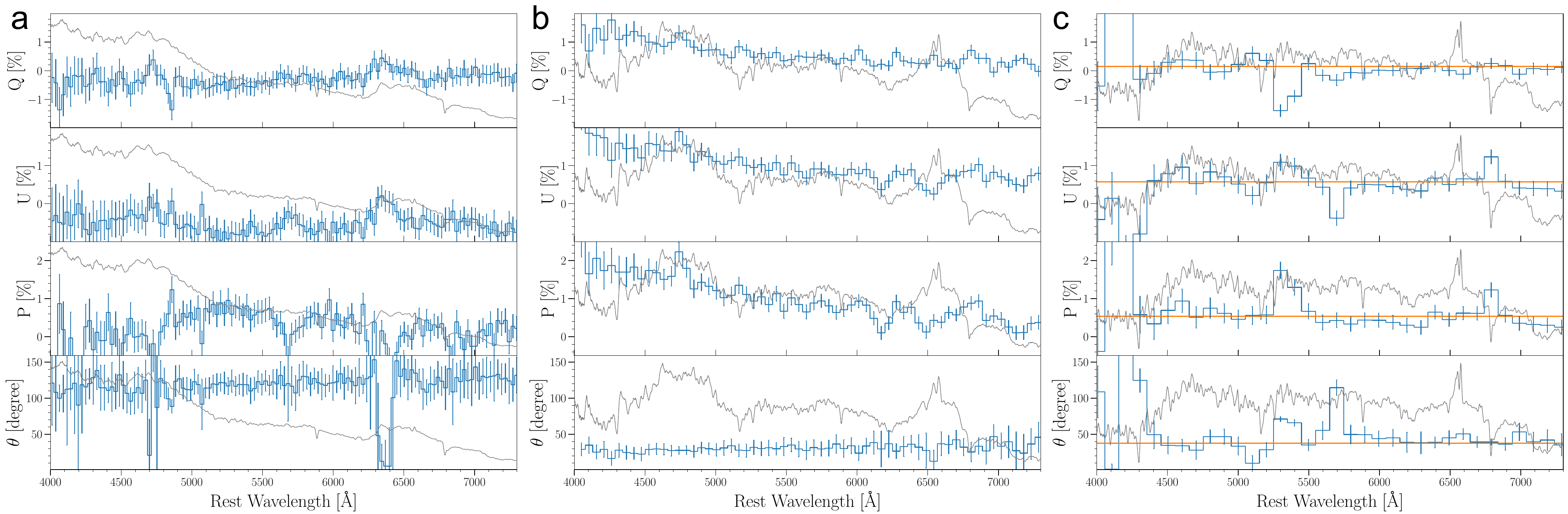}
\caption{
Evolution of spectropolarimetric properties of AT2023clx. Panel (a) shows the Stokes parameters of $Q$ (top panel) and $U$ (second panel), the polarization degree $P$ (third panel), and the polarization angle $\theta$ (bottom panel) on Day 5. The blue data point represents polarization values binned with $20$ pixels, corresponding to $\sim 26.8$ {\text \AA}. The error bars show the photon shot noise per bin. The underlying gray spectrum in each plot is the unbinned flux spectrum at the same epoch. AT2023clx exhibited the polarization degree of $\sim 0.6\%$ with a constant angle of $122.0 \pm 8.1$ degrees at the continuum component, without significant wavelength dependence.
Panel (b) is the same as panel (a), but for the data taken on Day 52. The polarization values are binned with $40$ pixels, corresponding to $\sim 53.6$ {\text \AA}. AT2023clx showed a wavelength-dependent polarization degree (0-2\%, increasing toward the blue) with a constant angle of $32.0 \pm 5.4$ degrees. The polarization data presented in panels (a) and (b) have been corrected for the host dilution effects (see Appendix \ref{appendix:correction}).
Panel (c) is the same as panel (a), but for the data taken on Day 121. The data are binned with $75$ pixels, corresponding to $\sim 100.5$ {\text \AA}. The wavelength-averaged polarization degree and angle, as represented by the orange lines, showed a constant degree of $\sim 0.5\%$ and a constant angle of $37.7 \pm 9.1$ degrees; this is interpreted to be originated from the background component (i.e., the static host component). 
}
\label{fig:polevolution}
\end{figure*}

Fig. \ref{fig:polevolution} shows the spectropolarimetric evolution. The continuum regions ($4900 {\text \AA} \lesssim \lambda \lesssim 6100 {\text \AA}$ and $6700 {\text \AA} \lesssim \lambda \lesssim 7300 {\text \AA}$) of the polarization spectrum at the first epoch ($+5.4$ days, hereafter Day 5) showed a constant polarization degree of $\sim 1\%$ with a constant polarization angle of $\sim 125$ degrees, while the polarization across several prominent features, in particular, the high-velocity components of Hydrogen and Helium, exhibit significant deviations from the continuum level. The second epoch of the polarization spectrum ($+52.1$ days, hereafter Day 52) was characterized by a rise in the polarization level toward shorter wavelengths, in particular from $\sim 0.2\%$ around $7000~\text{\AA}$ to $\sim 2.0\%$ around $4000~\text{\AA}$. The polarization angle displays a constant value of $\sim 35$ degrees across the observed wavelength range, which is different from the $\sim 125$ degrees observed on Day 5. In addition, no strong line polarization was discerned. At the third epoch ($\sim 121$ days, hereafter Day 121), the continuum polarization level decreased to $\sim 0.5\%$ while the polarization angle remains the same as observed on Day 52. The transient was sufficiently fainter than the host at such late phases, and thus we assume the data at Day 121 as extrinsic polarization (i.e., the static host polarization and the foreground polarization).

\begin{figure*}[t]
\epsscale{1.12}
\plotone{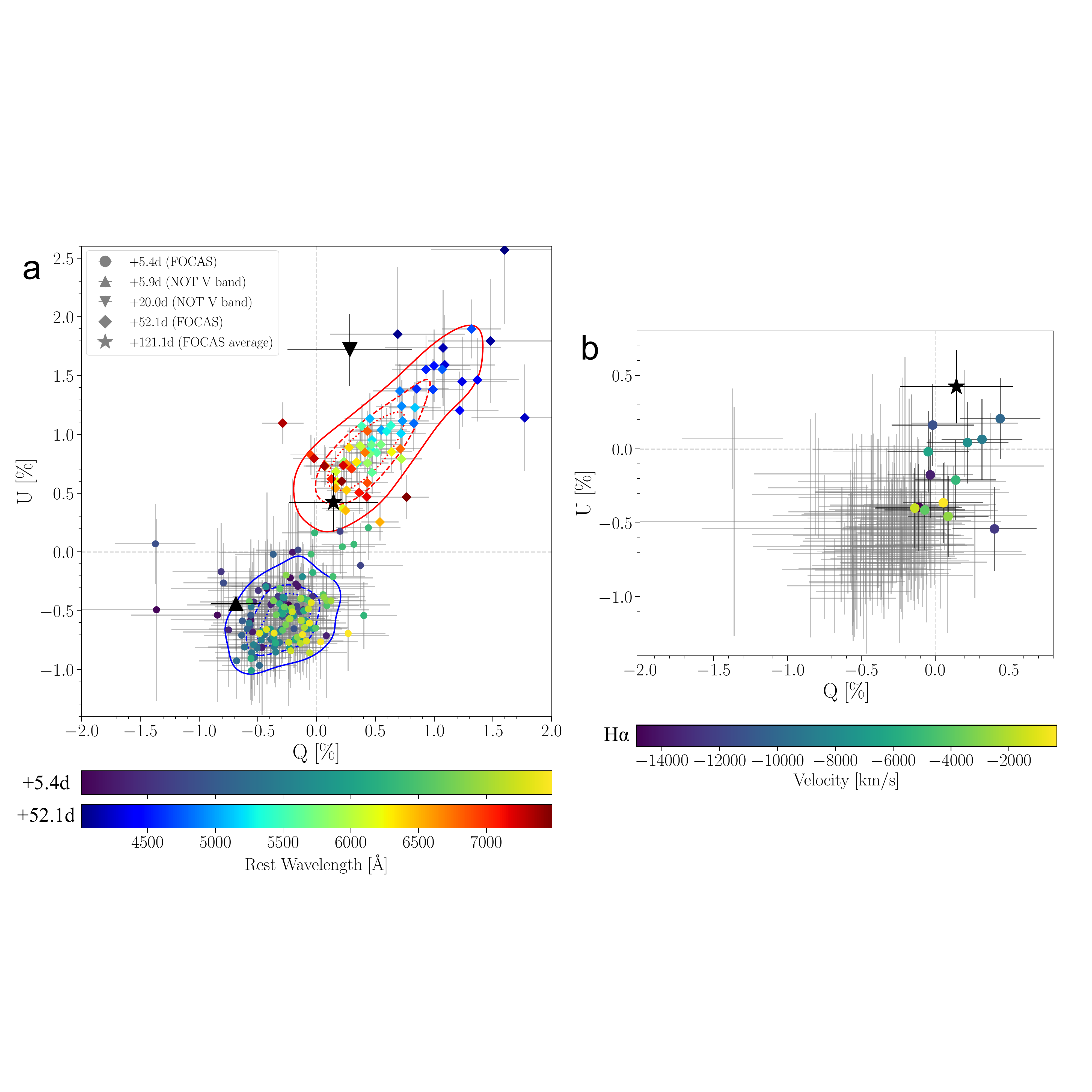}
\caption{
Left (Panel a):Time evolution of the polarization in the $Q$-$U$ diagram, including the measurements from the imaging polarimetry. To estimate the host dilution factor, we adopt the aperture size of $2^{\prime\prime}$ (see Appendix \ref{appendix:correction}). The color of each point corresponds to its wavelength, as indicated by the bottom colorbars; the viridis bar corresponds to the data on Day 5, while the rainbow bar corresponds to the data on Day 52. The black star symbol indicates the background polarization estimated from the averaged data on Day 121. Furthermore, the blue and red contours in the QU plane shows the 2D density distributions of the polarization data points on Day 5 and 52, respectively. The contours correspond to enclosed probability levels of 30\% (dotted), 50\% (dashed), and 80\% (solid), based on Gaussian kernel density estimation. The polarization angle flipped by $\sim 90$ degrees from Day 5 to 52 with respect to the background polarization.
Right (Panel b): The $Q$-$U$ diagram shown across the $\rm H\alpha$ profile on Day 5. The gray crosses are all the data points, and the color points are the data points for the $\rm H\alpha$, color-coded by the $\rm H\alpha$ velocity. The black star symbol indicates the background polarization estimated from the averaged data at Day 121. These figures show that the polarization degrees and angles at the wavelengths corresponding to the $\rm H\alpha$ deviate from the continuum region of the Day-5 data, indicating the line depolarization effect. The $Q$-$U$ position of the $\rm H\alpha$ is intermediate between the Day-5 continuum and the Day-121 data point, overlapping with the latter; this behavior supports our ISP estimation.
}
\label{fig:QUdiagram}
\end{figure*}

Fig. \ref{fig:QUdiagram} shows the polarization evolution on the $Q$-$U$ diagram, including the imaging polarimetry. The data on Day 5 are clustered around the point of $(Q, U) \sim (-0.20\% \pm 0.04\%, -0.40\%\pm 0.04\%)$, demonstrating no wavelength dependence. The data on Day 52 are distributed along a line connecting the Day-5 data and the origin of the $Q$-$U$ diagram, but on the opposite side with respect to the origin (i.e., 90-degree flip). The data in shorter wavelengths are more displaced from the origin. The contours help us to clearly separate the two datasets in a clean manner. The data on Day 121 display no wavelength dependence and overlap with the red end of the Day-52 data. We interpret that the spectropolarimetry on Day 121, in fact, measures the polarization zero point, i.e., the static polarization component unrelated to the transient. The imaging polarimetry data follow the similar trends seen in the spectropolarimetric data, supporting the analysis above. 

Comparing the polarization evolution of AT2023clx with those optical/UV TDEs for which polarization time series have been obtained \citep[e.g.,][]{Maund2020MNRAS, Leloudas2022NatAs, Patra2022MNRAS}, the other optical/UV TDEs kept roughly constant polarization degrees (without excess) and constant angles over $\sim 100$ days. On the other hand, AT2023clx showed different behaviors; the blue excess and the 90-degree polarization flip within the next $\sim 50$ days. The unique polarization behaviors observed in AT2023clx indicate the following points: (1) the system had an axis of symmetry, (2) the source that dominates the polarized emission has changed from Day 5 to 52, and (3) the persistent axial symmetry as inferred from a 90-degree flip of the polarization angle can be naturally explained by a prolate-to-oblate geometric transformation with the same symmetry axis (for details, see Sec. \ref{sec:discussion}).

\section{Discussion} \label{sec:discussion}

\subsection{Geometrical Alignment behind The Transient}

 The unprecedented polarization time series of the TDE AT2023clx implies the common axial symmetry shared by the emission sources of the TDE and the underlying host nucleus. On Day 5, AT2023clx was brighter than the host galaxy. Indeed, the early-phase spectra displayed the transient-dominated features, and the polarization should also be dominated by the transient emission. The detected polarization degree ($\sim 1\%$) is incompatible with the picture that the transient light was mainly emitted through the shock collision between different parts of a tidally-disrupted tail \citep{Piran2015ApJ, Lu2020MNRAS, Steinberg2024Nature}, since this process predicts much stronger polarization signals ($\gtrsim 10\%$) as detected in a TDE previously \citep{Liodakis2023Sci}. This favors another mechanism, namely an optically-thick outflow driven by accretion flow onto the central SMBH \citep{Strubbe2009MNRAS, Metzger2016MNRAS, Leloudas2022NatAs} as the probable emission source in the case of AT2023clx. The outflow likely deviates from spherical symmetry with a specific direction, which is compatible with the polarization on Day 5 \citep{Hoflich1991AA, Dai2018ApJ, Thomsen2022ApJ}. Note that this early-time polarization is also different from that of AT2019qiz, which showed negligible polarization at peak brightness, suggesting the presence of a nearly spherical reprocessing layer \citep{Patra2022MNRAS}. In addition, the polarization degrees and angles at the strong emission lines (H$\alpha$ and H$\beta$) deviated from those of the continuum (see also Fig.\ref{fig:QUdiagram}b). Analogous to supernovae \citep[e.g.,]{Leonard2000ApJ, Patat2011AA, Yang2023MNRAS, Uno2023ApJ}, it is likely that this was due to the line depolarization effect caused by the recombination, and the polarization properties at these strong emission lines approached those of the static background/foreground component, including the interstellar polarization. This is indeed confirmed by the Day-121 data that overlapped with the Day-5 data points at the strong emission lines on the $Q$-$U$ diagram, defining the intrinsic zero point in the $Q$-$U$ plane (see below).

Toward Day 52, the electron scattering within the outflow should become ineffective as the transient light fades away and the optical depth decreases. If the transient is embedded in a dust-rich environment, the secondary source of the polarized dust scattering may become progressively dominant over time. The bright transient light illuminated the surrounding dust, producing a scattering light echo that can be highly polarized at large scattering angles and deviate from spherical symmetry. The emission of such a scattered light echo can be described by convolving the light curve of the transient with a dust kernel that characterizes the geometric and scattering properties of the dust cloud \citep[see, e.g.,][]{Chevalier1986ApJ, Patat2005MNRAS, Yang2017ApJ}. The key prediction in this scenario is stronger polarization toward the blue once the echo dominates the polarization signal, due to a combined effect of the blue color of the illuminating transient light at its peak luminosity and the preference of the dust scattering toward the shorter wavelength \citep{Nagao2018ApJ,Nagao2018MNRAS}. This explains the polarization with the blue excess, as was observed on Day 52. This scenario is also supported by the weak near-infrared (NIR) excess that emerged in the late phase \citep[$L_{\rm NIR} \sim 4\times 10^{40} {\rm ~ erg ~ s^{-1}}$,][]{Charalampopoulos2024AA}, demonstrating the existence of dust close to the SMBH. The NIR excess may arise from dust re-radiation, with a luminosity comparable in order of magnitude to that from dust scattering via light echoes ($L_{\rm NIR} \sim L_{\rm sc}$). Then, the polarization degree by the scattering ($P_{\rm sc}$) is determined by the fraction of the (delayed) scattered photons to the transient's photons directly reaching to the observer: $P_{\rm sc} \lesssim L_{\rm sc}/L_{\rm bol} \sim 8\%$, where $L_{\rm bol} \sim 5\times 10^{41} {\rm ~ erg ~ s^{-1}}$ is the bolometric luminosity around Day 52 \citep{Zhu2023ApJL, Hoogendam2024MNRAS,Charalampopoulos2024AA}. Although the estimated polarization is actually a maximum value due to the cancellation of polarization by multiple scattering and viewing angle effects, the NIR excess and the polarization on Day 52 are roughly consistent.

\begin{figure}[t]
\epsscale{1.12}
\plotone{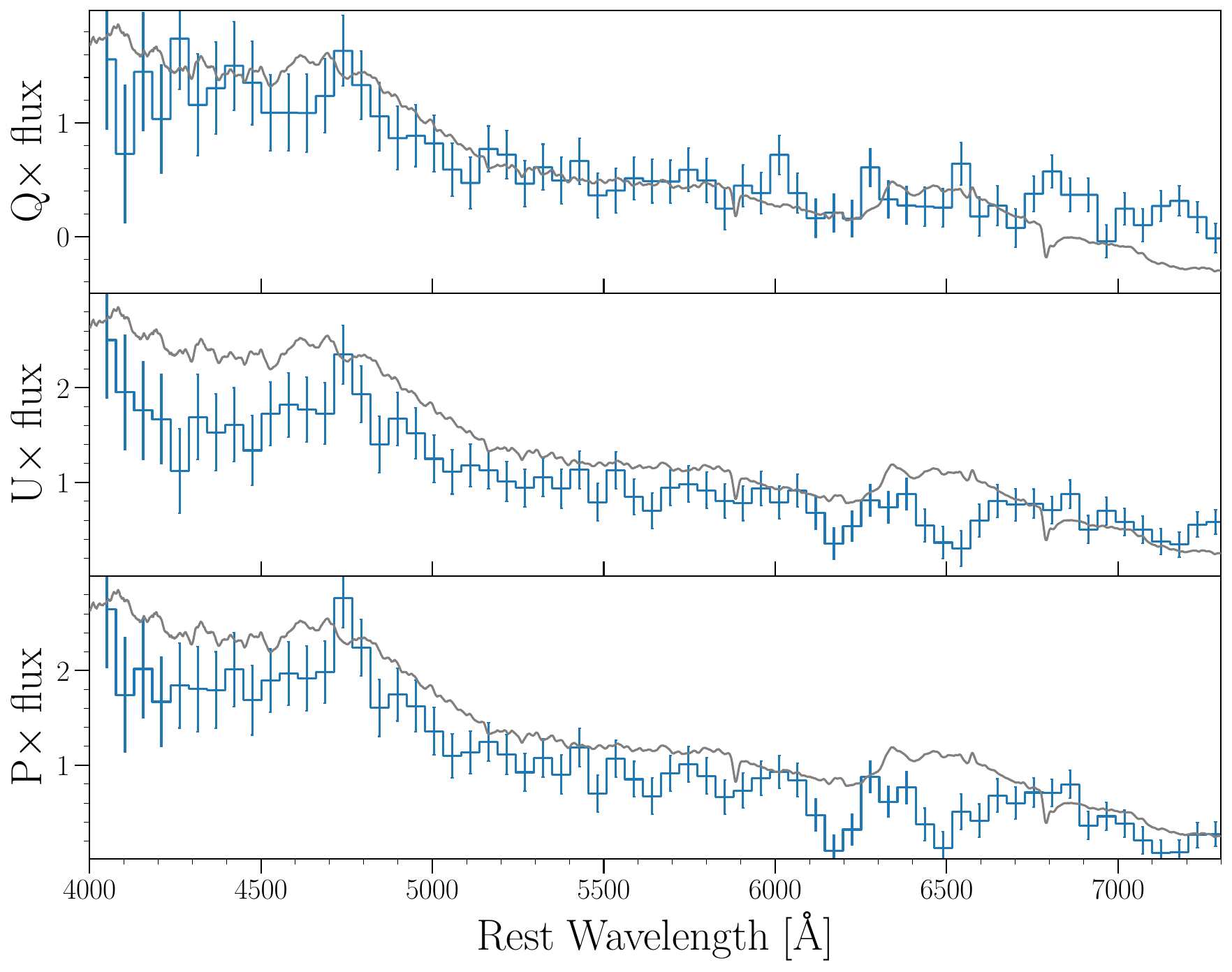}
\caption{
Polarized fluxes of AT2023clx on Day 52. The polarized flux is calculated by $Q[\%] \times$, $U[\%] \times$, and $P[\%] \times$ flux obtained on Day 52, where the flux is normalized by the mean flux over the spectral wavelengths. The underlying gray spectrum in each panel is the unbinned spectrum on Day 5. The polarized fluxes correlated with the spectrum on Day 5 demonstrate that the polarization on Day 52 is originated from light echoes.}
\label{fig:polflux}
\end{figure}

Fig. \ref{fig:polflux} compares the observed flux on Day 5 and the polarized flux on Day 52. The polarized flux roughly follows the spectral shape of the Day-5 flux, clearly demonstrating that the Day-52 polarization originates from light echoes. Moreover, this figure provides an insight into the dust properties. Polarization from dust scattering generally reflects wavelength-dependent scattering efficiency, i.e., opacity, determined by dust grain size. This is one of the main causes of dust polarization features with the bluer excess. However, the polarized flux on Day 52 closely traces the continuum shape of the Day-5 flux, showing a mild blue excess. This suggests that the dust opacity producing the polarization has a weak wavelength dependence, possibly implying that the nuclear dust in AT2023clx consists of relatively large grains ($\gtrsim$ micron size). This not only supports our interpretation of the scattering-polarization scenario, but also possibly offers helpful insights into the properties of nuclear dust.

Furthermore, the scattering-polarization scenario can provide an explanation for another striking feature in the second phase, namely the angle flip of $\sim 90$ degrees from Day 5 to 52, indicating a prolate-to-oblate transition in the geometry of the emitting regions. Therefore, we suggest a simple hypothesis that the axis of the transient outflow and the dust plane were oriented perpendicularly. In general, the polarization degree reflects the radiation mechanism and asymmetry of the system, while the polarization angle traces the direction of the bulk structure of the polarization source. Since scattered light acquires linear polarization perpendicular to the scattering plane, the 90-degree flip indicates that the outflow (from Day 5) and the dust plane (from Day 52) should intersect in the orthogonal directions at least on the two-dimensional projection onto the sky. The most straightforward configuration is that they are also perpendicular in three-dimensional space.

Eventually, as indicated by the light curves of AT2023clx \citep{Hoogendam2024MNRAS, Charalampopoulos2024AA} on Day 121, both the emission along the direct line of sight and the scattered light echo have faded substantially. On Day 121, the blue excess in the polarization spectrum can no longer be identified, and the measurement presents the static polarization (i.e., foreground interstellar polarization including the Milky Way contribution, plus polarization originated from the static host light). On the $Q$-$U$ diagram, this static polarization overlaps with the red end of the Day-52 data, indicating that the latter is a superposition of a scattering polarization and a static polarization. These two components share the same axial symmetry (i.e., the system of the accretion disk and dust plane). We note that the static polarization as defined by the polarization on Day 121 is consistent with that seen at the strong emission lines in the Day-5 data, reinforcing the robustness of the scenario presented here (see also Fig. \ref{fig:QUdiagram}b).

\subsection{Possible Origins}

The spectropolarimetric time series of AT2023clx, indicating that the outflow is perpendicularly aligned to the dust plane, is a unique finding that has never been detected in previous TDEs. Since the timescale is too short to generate newly-formed dust, the blue-excess polarization might originate from pre-existing dust. Furthermore, in the standard TDE picture, the angle of the star plunging into the central BH is expected to be random. The associated outflow, as driven through the accreting stellar debris, is then likely oriented to a random direction as well, since the outflow direction is expected to be determined by the global structure of the accreting debris that should follow the stellar orbit of the disrupted star in its direction, rather than the local structure linked to BH parameters \citep[e.g., BH spin,][]{Liska2021MNRAS}. Therefore, there is no reason to expect the correlation between the direction of the TDE outflow and that of the pre-existing dust; the chance probability for having the polarization angle within $\sim 90 \pm 5$ degrees is less than $10\%$ (see Appendix \ref{appendix:chance_probability}), which potentially requires a natural mechanism to align the stellar orbit with the dust plane. 

 One natural interpretation is that the disrupted star approached the SMBH through a nuclear dusty region moving around the SMBH. Assuming that the dense dusty region captures nuclear stars or drives star formation within the cloud, the stellar orbit along the dusty region can be naturally explained, leading to the perpendicular alignment between the outflow driven by the TDE and the dusty plane. In this context, an interesting possibility is that the LINER nature of the host galaxy core might play a key role. Although the nature of LINERs has not been completely understood, they are likely placed as transitional objects bridging non-AGNs and AGNs \citep{Ho2001ApJL, Satyapal2004AA, Satyapal2005ApJ, Ho2009ApJ}. If LINERs have scaled-down structures similar to AGNs, such as a small dust torus and an accretion disk, stars might be captured by the accretion disk or formed within the torus region, leading to their plunging angles into the SMBH aligning with the disk/torus plane. Indeed, it has been theoretically proposed that an AGN accretion disk might affect the dynamics of stars in a nuclear region \citep{Artymowicz1993ApJ, Fabj2020MNRAS}, enhancing the intrinsic event rate of TDEs associated with AGNs \citep{Wang2024ApJ}. AT2023clx may then demonstrate the existence of the disk-captured TDE. 

Alternatively, AT2023clx might be an AGN-like activity, potentially linked to changing-look AGNs \citep[CLAGNs;][for review]{LaMassa2015ApJ,Ricci2023NatAs} or AGN major flares \citep{Graham2017MNRAS, Graham2020MNRAS}, rather than a TDE. More broadly, AT2023clx possibly provides insights into ANTs with characteristics overlapping between TDEs and AGN-related flares. This scenario may raise a question of why an AGN-like activity would look like a TDE in observational properties, but we note that the observational appearance of the outburst is mainly determined by how much and how quickly mass accretes onto the SMBH \citep{Uno2020ApJ, Matsumoto2021MNRAS}. Based on the faint nature of AT2023clx as a TDE and an analogy with the previously-analyzed faint TDEs \citep[e.g., iPTF16fnl and AT2020wey,][]{Blagorodnova2017ApJ, Mockler2019ApJ, Charalampopoulos2023AA}, it would require a mass accretion of $\sim 0.1 {\rm ~M_{\odot}}$ with a timescale of $\sim 100$ days; this is marginally super-Eddington accretion, which is within the range derived for some AGNs \citep{Du2015}. One previous example might support the AGN-like scenario; in TDE Arp299-B AT1, a spatially-resolved and expanding radio jet was found to be misaligned with the pre-existing AGN torus \citep{Mattila2018Sci}. This was used as a strong argument for this event to have originated from a TDE rather than an AGN-triggered activity. The same argument would apply to AT2023clx, but to conclude the opposite case, i.e., it might be an AGN-triggered activity.

Although these scenarios can naturally explain the features of AT2023clx, challenges also remain. For the TDE scenario (including a possible effect of the `AGN' torus around AT2023clx), the weak NIR excess and dust polarization favor a relatively clearer environment than standard AGNs, raising doubts about whether a sufficiently dense region/cloud exists to influence stellar orbits or star formation. For the AGN-activity scenario, if a sufficiently massive torus/disk could be developed in the LINER case can be a problem in this scenario; the mechanism driving rapid mass accretion, even for CLAGNs, has been still unknown. Further, showing TDE-like observational properties in the AGN-activity scenario is largely speculative. Further theoretical study, as well as additional observational examples, will be required to discriminate the origin of nuclear transients.

\section{Conclusions} \label{sec:conclusions}

 We present three-epoch spectropolarimetry of a nearby nuclear transient AT2023clx, spanning from Day 5 to Day 121. On Day 5, we detected a high constant polarization of $\sim 1\%$ with a constant polarization angle of $\sim 125$ degrees, likely originated from the optically-thick transient outflow. On Day 52, we detected a blue-excess polarization reaching $\sim 2\%$ with a constant angle of $\sim 35$ degrees, indicating that the polarized emission from a scattered light echo has become significant on Day 52. On Day 121, our observation shows a wavelength-independent polarization spectrum at a level of $\sim$0.5 and a polarization angle of $\sim 35$ degrees, corresponding to static polarization from the host galaxy. 

The 90-degree polarization flip between the first two epochs implies that the scattering sources of the two epochs are perpendicularly aligned, namely the direction of the transient outflow is perpendicular to the dust plane. One natural explanation yields the features is stellar capture or star formation within a dense dusty cloud in the nuclear region. An interesting possibility is that such a dusty region may be linked to a dusty torus around the SMBH, given that the host core is classified as LINER. An even more speculative possibility is that the event was triggered by an AGN-like activity, potentially linked to CLAGNs or ANTs.

Without polarimetric observations, we would have missed the unique geometrical properties of AT2023clx. Our results underscore the diagnostic power of spectropolarimetry in probing the geometry and physical origin of nuclear transients. In particular, time-series spectropolarimetric observations -- spanning from the early to late phases -- can reveal evolving polarization components that are inaccessible in single-epoch data. Expanding such observations to TDEs with a broader range of luminosities will provide a unique avenue to explore the diversity of transient phenomena and clarify the connections among TDEs, AGN-like events, and ANTs. AT2023clx potentially provides valuable insights to map the observational classes of nuclear transients to different origins. Our findings indicate that some of the previously-identified `TDEs' may fall into other classes and vice versa, which may further address the question of the increased TDE rate in AGNs. Furthermore, the polarized flux might suggest a weak wavelength dependence of the dust opacity, implying that the nuclear region around AT2023clx may favor relatively large dust grains. This highlights the potential of spectropolarimetry not only to probe the properties of nuclear transients, but also to investigate the properties of nuclear environments using the transients.

\begin{acknowledgments}
This research is based on observations obtained at the Subaru Telescope (S23A-052, PI: K.U.) operated by the National Astronomical Observatory of Japan (NAOJ), the Seimei Telescope at the Okayama observatory of Kyoto University (23A-K-0001, PI: K.M.), the Nordic Optical Telescope (66-019, PI: P.C.), and e-MERLIN (DD16004, PI: M.P.-T.). The Seimei telescope is jointly operated by Kyoto University and the NAOJ, with assistance provided by the Optical and Near-Infrared Astronomy Inter-University Cooperation Program. The spectra taken by the Seimei telescope are obtained with the KASTOR (Kanata And Seimei Transient Observation Regime) campaign. The authors thank Mitsuru Kokubo, Takeo Minezaki, and Jason T. Hinkle for valuable discussions. K.U. acknowledges financial support from Grant-in-Aid for the Japan Society for the Promotion of Science (JSPS) Fellows (22J22705 and 22KJ1986), and support from the Hayakawa Satio Fund awarded by the Astronomical Society of Japan. K.M. acknowledges support from JSPS KAKENHI grant Nos. JP20H00174, JP20H04737, JP24H01810, and 24KK0070, and support from the JSPS Open Partnership Bilateral Joint Research Projects between Japan and Finland (JPJSBP120229923). T.N. acknowledges support from the Research Council of Finland projects 324504, 328898, and 353019. G.L. was supported by a research grant (19054) from VILLUM FONDEN. S.M. was funded by the Academy of Finland project 350458. J.M. and M.P.-T. acknowledge financial support from the grant CEX2021-001131-S funded by MICIN/AEI/ 10.13039/501100011033. J.M. acknowledges financial support from the grant PID2021-123930OB-C21 funded by MICIN/AEI/ 10.13039/501100011033, by `ERDF A way of making Europe' and by the `European Union'. M.P.-T. acknowledges financial support through grant PID2020-117404GB-C21, funded by MCIN/AEI/ 10.13039/501100011033. J.M. acknowledges the Spanish Prototype of an SRC (SPSRC) service and support funded by the Spanish Ministry of Science, Innovation and Universities, by the Regional Government of Andalusia, by the European Regional Development Funds and by the European Union NextGenerationEU/PRTR. The SPSRC acknowledges financial support from the State Agency for Research of the Spanish MCIU through the `Center of Excellence Severo Ochoa' award to the Instituto de Astrofísica de Andalucía (SEV-2017-0709) and from the grant CEX2021-001131-S funded by MICIN/AEI/ 10.13039/501100011033. e-MERLIN is a National Facility operated by the University of Manchester at Jodrell Bank Observatory on behalf of STFC. M.P. acknowledges support from a UK Research and Innovation Fellowship (MR/T020784/1). The data within this paper will be available on the Weizmann Interactive Supernova Data Repository (WISeREP), or upon reasonable request to the corresponding authors.
\end{acknowledgments}

\begin{contribution}
K.U., K.M., and T.N. initiated the project. K.U. led the spectral polarimetry data analysis, discussion, and manuscript preparation. K.M. and T.N. organized the efforts for the interpretation of the results and assisted in the manuscript preparation. K.M. also organized the effort for the rapid spectral classification of nuclear transients with the Seimei telescope. G.L., P.C., and S.M. provide valuable insights into the discussion and interpretation. P.C. obtained the imaging polarimetry data with the Nordic Optical telescope, and T.N. analyzed the data. K.A. supported in planning of the observation strategy with the Subaru as well as analyzed the pre-outburst AGN spectral classification. K.T. performed the Seimei observation, and analyzed the first classification spectrum. M.K. developed a web interface to search for TDE candidates from public transient-discovery alerts, which assisted in the selection of AT2023clx as a TDE candidate. M.P.-T. led the e-MERLIN radio proposal. J.M. reduced and analyzed the e-MERLIN data. M.P. summarized the host spectrum data taken by the VLT. T.R. provided NIR data of AT2023clx. All authors contributed to the discussion.
\end{contribution}

\facilities{Subaru, Seimei, NOT, e-MERLIN}
\software{IRAF, Astropy \citep{astropy13,astropy18}, CASA \citep{CASA2022PASP}}

\appendix

\section{Radio Follow-up Observations} \label{appendix:radio}

Five days after the discovery, a radio point source was detected at the position consistent with AT2023clx. The flux density was $0.40 \pm 0.08 {\rm ~mJy}$ at $15.5 {\rm ~GHz}$ \citep{Sfaradi2023TNSAN}. We also performed radio follow-up observations with the enhanced Multi Element Remotely Linked Interferometer Network (e-MERLIN), i.e., a very large baseline interferometer in Europe. We conducted four observing runs on January 13, 15, 16, and February 4, 2024, at a central frequency of $5.1 {\rm ~GHz}$ containing 4 spectral windows uniformly distributed with a bandwidth of $512 {\rm ~MHz}$. The main calibration was performed using the e-MERLIN CASA pipeline version v1.1.19 \citep{Moldon2021ascl} using CASA version 5.8 \citep{CASA2022PASP}. Additional manual flagging, the amplitude and phase self-calibration using three sources in the field were performed using wsclean \citep{Offringa2014MNRAS} and CASA. The phase reference calibrator was J$1157+1638$, correlated at position $\rm{RA(J2000.0)} = 11^{\rm h}57^{\rm m}34^{\rm s}.836$ and ${\rm Dec(J2000.0)} = 16^{\circ}38^{\prime}59^{\prime\prime}.65$. Target scans were 6 min long and were interleaved with 2 min scans on the phase calibrator. 3C286 and OQ208 were also observed as the flux and bandpass calibrators, respectively. 

No radio source was detected in the individual e-MERLIN runs within $8^{\prime\prime}$ from the optical position. We further combined the multi-epoch data to form a single deep image. The resulting synthesized beam was $98 {\rm mas} \times 42 {\rm mas}$. We derived the detection upper limit of the final image as $< 36 {\rm ~\mu Jy}$ at 3-sigma level. AT2023clx was not a radio-loud event; we did not detect a radio counterpart, except for a single weak detection at the early phase. Some TDEs show late-time radio rebrightening \citep[e.g.,][]{Alexander2016ApJ}, and thus continuous radio follow-up observations possibly play a helpful role in distinguishing different scenarios for AT2023clx proposed in the present work.

\section{Data Correction for Spectropolarimetry} \label{appendix:correction}

To correct the spectropolarimetric data, we obtained standard calibration data: highly polarized and unpolarized standard stars, and fully-polarized flat lamp data. First, using the unpolarized standard stars, we calibrated the instrumental polarization of FOCAS. We further calibrated the offset of the polarization angle from the reference axis on the celestial plane using the strongly polarized standard stars. Then, we calibrated the wavelength dependence of the polarization angle using the fully-polarized flat lamp data. For flux calibration, we also obtained spectroscopic standard stars with the same polarization setup.

Furthermore, to evaluate the intrinsic polarization of the nuclear transient, we used two additional polarization corrections: the bias correction and the host contribution correction. For the bias correction, we used the standard method described in \citet{Wang1997ApJL} as follows:
\begin{equation}
  P_{\rm true} = P_{\rm obs} - \sigma_{P},
\end{equation}
where $P_{\rm true}$ and $P_{\rm obs}$ are polarization degrees after and before the bias correction, respectively, and $\sigma_{P}$ is defined as follows: $\sigma_{P} = \sigma^{2} / P_{\rm obs}$, where $\sigma$ is the error of the polarization degree. In the original mathematical definition ($P = \sqrt{Q^{2} + U^{2}}$), the polarization degrees must be positive numbers, which creates artificial polarization. Using this calibration, the artificial polarization degrees were corrected to zero. 

To correct the effect of light dilution by the host galaxy contribution, we applied the host contribution correction following \citet{Leloudas2022NatAs}. In this method, the observed polarization including the host polarization ($Q_{\rm obs}$, $U_{\rm obs}$) is described as follows:
\begin{equation}
  Q_{\rm obs} = \frac{Q_{\rm transient} I_{\rm transient} + Q_{\rm host} I_{\rm host}}{I_{\rm transient} + I_{\rm host}}, 
\end{equation}
where $Q_{\rm transient}$ and $Q_{\rm host}$ are the intrinsic polarization originated from the nuclear transient and the host galaxy, respectively, and $I_{\rm transient}$ and $I_{\rm host}$ are the fluxes from the transient and host galaxy, respectively. Here, we defined the host contribution ratio ($\alpha$) as follows: $\alpha(\lambda) = I_{\rm host}(\lambda)/I_{\rm obs}(\lambda)$, where $I_{\rm obs} = I_{\rm transient} + I_{\rm host}$ is the observed flux including the host galaxy. Assuming the host polarization was zero ($Q_{\rm host} \approx 0$), we derived the intrinsic polarization of the transient as follows: $Q_{\rm transient}(\lambda) = Q_{\rm obs}(\lambda) / (1 - \alpha(\lambda))$. We also applied the same calibration for $U_{\rm obs}$. 

For Day 5, we had no photometry data on the same day. Alternatively, we used the photometry data obtained one day before ($+4.4$ days) by the Seimei telescope. We performed aperture photometry adopting the aperture size ($\phi$) of $\sim 2^{\prime\prime}$, and then the apparent magnitude including the host galaxy was estimated to be $16.3$ mag in the $g$ band. Note that the point spread function size of the Seimei imaging data was estimated to be $\sim 2^{\prime\prime}$, although the seeing size of the spectropolarimetric data on Day 5 was $1^{\prime\prime}$ with the slit width set to $1^{\prime\prime}$. Using the apparent magnitude, we calibrated the flux level of the spectrum, and obtained $I_{\rm obs}$. 

The host magnitude was estimated from the pre-outburst spectra taken by the SDSS, as well as by the Multi Unit Spectroscopic Explorer \citep[MUSE;][]{Bacon2010SPIE} mounted on the Very Large Telescope\footnote{\url{http://archive.eso.org/dataset/ADP.2021-03-11T09:40:12.294}}. Note that the fiber size of the SDSS spectrum was $3^{\prime\prime}$ centered on the core, while one MUSE pixel size was $0.2^{\prime\prime}$. We combined the MUSE IFU spectral cubes within $\sim 2^{\prime\prime}$, i.e., the same aperture size used for the photometry, and reconstructed a flux-calibrated host spectrum. Then, we scaled the SDSS spectrum to match the flux level of the flux-calibrated MUSE spectrum and obtained $I_{\rm host}$, because the spectral range of the MUSE spectra did not cover that of our spectropolarimetric data. 

For Day 52, we also performed the imaging observation with FOCAS, and we estimated the magnitude including the host galaxy from the FOCAS imaging data on the same day. To align the host contribution effect with Day 5, the aperture size was set to $\phi \sim 2^{\prime\prime}$. The apparent magnitude including the host galaxy was estimated to be $16.7$ mag in the $V$ band. The host flux was estimated with the same method as Day 5. To test the effects of the aperture size, we also performed the host contribution correction with $\phi \sim 1^{\prime\prime}$ on Day 52, corresponding to the seeing size and the slit width in the FOCAS observation on Day 52. Adopting the large aperture size, the host contribution increased, but the trend of the polarization behaviors did not change. Thus, we concluded that the polarization analysis was robust and insensitive to the host contribution. Note that we did not perform the host contribution correction for the Day-121 data, because we assumed that the outburst did not contribute to the polarization in this epoch and that the polarization might well be dominated by the host component. This assumption was also supported by line depolarization at the Day-5 data. 

To further investigate the host contribution effects, we tested the case of $Q_{\rm host} \neq 0$. In this case, the host correction equation is described as follows:
\begin{equation}
  Q_{\rm transient} = \frac{1}{1 - \alpha(\lambda)} Q_{\rm obs} - \frac{\alpha(\lambda)}{1 - \alpha(\lambda)} Q_{\rm host}.
\end{equation}
This correction corresponds to the host polarization subtraction. Using the averaged polarization value of the Day-121 data as the host polarization, we obtained the intrinsic polarization originated from the transient. The intrinsic polarization showed the angle flip of 90 degrees, supporting the robustness of our interpretation and discussion.

\section{Chance probability for the orbital alignment} \label{appendix:chance_probability}

Assuming the completely random stellar motion in TDEs, the chance probability of the alignment of the plaguing angle is determined by the solid angle. The polarization angles of Day 5, 52, and 121, were estimated as $\sim 122$, $\sim 32$, and $\sim 37$ degrees, respectively. These values are aligned within a range of $90 \pm 5$ degree, i.e., the acceptable range in the deviation of the half-opening angle ($\theta$) from the perfect alignment ($90$ degree) is only $5$ degree. Given the solid angle ($\Omega$) described as $\Omega = 4\pi \sin \theta$, the chance probability ($P$) that this alignment was realized from a random distribution is $P \approx \Omega / 4\pi = 0.087$. Therefore, the chance probability for aligning the polarization angles within $\sim 90 \pm 5$ degrees based on the random stellar encounter to an SMBH is $\sim 8.7\%$. 

\bibliography{manuscript}{}
\bibliographystyle{aasjournal}

\end{document}